\documentclass[aps,longbibliography,
  twocolumn,amsmath,
  amssymb,nofootinbib, superscriptaddress, preprintnumbers]{revtex4-1}

\usepackage{graphics}      
\usepackage{graphicx}      
\usepackage{longtable}     
\usepackage{url}           
\usepackage{bm}            
\usepackage{xcolor}
\usepackage{mathrsfs}
\usepackage[colorlinks,allcolors=blue]{hyperref}
\usepackage{blindtext}
\usepackage{hyperref}
\usepackage{float}
\usepackage{bbold}
\usepackage{lipsum}
\usepackage[normalem]{ulem}
\usepackage{orcidlink}
\usepackage{academicons}
\usepackage{xcolor}
\usepackage{subfigure}
\usepackage{booktabs}
\usepackage{hyperref}

\newcommand{\orcid}[1]{\href{https://orcid.org/#1}{\textcolor[HTML]{A6CE39}{\aiOrcid}}}

\begin{document}
\preprint{MPP-2024-12}

\title{Application of  Neural Networks for
the Reconstruction of Supernova Neutrino Energy Spectra Following Fast Neutrino Flavor Conversions}


\newcommand*{\MPP}{\textit{\small{Max-Planck-Institut f\"ur Physik (Werner-Heisenberg-Institut), Boltzmannstr. 8, 85748 Garching, Germany}}}

\author{Sajad Abbar \orcidlink{0000-0001-8276-997X}   } 
\affiliation{\MPP}
\author{Meng-Ru Wu \orcidlink{0000-0003-4960-8706}}
\affiliation{Institute of Physics, Academia Sinica, Taipei, 11529, Taiwan}
\affiliation{Institute of Astronomy and Astrophysics, Academia Sinica, Taipei, 10617, Taiwan}
\affiliation{Physics Division, National Center for Theoretical Sciences, Taipei 10617, Taiwan}

\author{Zewei Xiong \orcidlink{0000-0002-2385-6771} }
\affiliation{GSI Helmholtzzentrum {f\"ur} Schwerionenforschung, Planckstra{\ss}e 1, D-64291 Darmstadt, Germany}

\begin{abstract}

Neutrinos can undergo fast flavor conversions (FFCs) within extremely dense astrophysical environments such as core-collapse supernovae (CCSNe) and neutron star mergers (NSMs). In this study, we explore FFCs in a \emph{multi-energy} neutrino gas, revealing that when the FFC growth rate significantly exceeds that of the vacuum Hamiltonian, all neutrinos (regardless of energy) share a common survival probability dictated by the energy-integrated neutrino  spectrum.
We then employ physics-informed neural networks (PINNs) to predict the asymptotic outcomes of FFCs within such a multi-energy neutrino gas. These predictions are based on the first two moments of neutrino angular distributions for each energy bin, typically available in state-of-the-art CCSN and NSM simulations.
Our PINNs  achieve errors as low as  $\lesssim6\%$ and $\lesssim 18\%$ for predicting the number of neutrinos in the electron channel and the relative absolute error in the neutrino moments, respectively. 

 \end{abstract}

\maketitle

\section{Introduction}

Core-collapse supernovae (CCSNe) and neutron star mergers (NSMs) represent some of the most extreme astrophysical settings, where neutrino emission plays a crucial role~\cite{Burrows:2020qrp, Janka:2012wk, Foucart:2022bth, Kyutoku:2021icp, Colgate:1966ax, Lattimer:1974slx, mirizzi2016supernova}. Within the dense and extreme conditions of these environments, neutrinos undergo a complex flavor conversion process triggered by their coherent forward scatterings with the dense background neutrino gas~\cite{pantaleone:1992eq, sigl1993general, Pastor:2002we,duan:2006an, duan:2006jv, duan:2010bg, Mirizzi:2015eza, volpe2023neutrinos}.

One of the latest advancements in this field involves the discovery of \emph{fast} flavor conversions (FFCs), which can take place on scales significantly shorter than those anticipated in the vacuum~\cite{Sawyer:2005jk, Sawyer:2015dsa,
Chakraborty:2016lct, Izaguirre:2016gsx,Capozzi:2017gqd, Wu:2017qpc, Wu:2017drk, Abbar:2017pkh, Abbar:2018beu,Capozzi:2018clo, 
Martin:2019gxb, Abbar:2018shq, Abbar:2019zoq, Capozzi:2019lso, George:2020veu, Johns:2019izj, Martin:2021xyl, Tamborra:2020cul,  Sigl:2021tmj,   Morinaga:2021vmc, Nagakura:2021hyb,  Sasaki:2021zld, Padilla-Gay:2021haz, 
Xiong:2021dex, Capozzi:2020kge, Abbar:2020qpi, Capozzi:2020syn, DelfanAzari:2019epo, Harada:2021ata,  Abbar:2021lmm, Just:2022flt, 
Padilla-Gay:2022wck, Capozzi:2022dtr, Zaizen:2022cik, Shalgar:2022rjj,  Kato:2022vsu, Zaizen:2022cik,  Bhattacharyya:2020jpj, Wu:2021uvt, Richers:2021nbx, Richers:2021xtf, Dasgupta:2021gfs, Capozzi:2022dtr, Nagakura:2022kic, Ehring:2023lcd, Ehring:2023abs, Xiong:2023upa, Fiorillo:2023hlk, Nagakura:2023wbf, Martin:2023gbo, Fiorillo:2023mze,
Grohs:2023pgq}.
A condition both necessary and sufficient for the occurrence of FFCs is that the angular distribution of the neutrino lepton number, defined as,
\begin{equation}
\begin{split}
  G(\mathbf{v}) =
  \sqrt2 G_{\mathrm{F}}
  \int_0^\infty  \frac{E_\nu^2 \mathrm{d} E_\nu}{(2\pi)^3}
        &[\big( f_{\nu_e}(\mathbf{p}) -  f_{\nu_x}(\mathbf{p}) \big)\\
              &- \big( f_{\bar\nu_e}(\mathbf{p}) -  f_{\bar\nu_x}(\mathbf{p}) \big)],
 \label{Eq:G}
\end{split}
\end{equation}
crosses zero at some $\mathbf{v} = \mathbf{v}(\mu,\phi_\nu)$, with $\mu =\cos\theta_\nu$~\cite{Morinaga:2021vmc}.
Here, $G_{\rm{F}}$ represents the Fermi coupling constant, $E_\nu$, $\theta_\nu$, and $\phi_\nu$ are the neutrino energy,  
the zenith, and azimuthal angles of the neutrino velocity, respectively. 
The $f_{\nu}$'s are the neutrino 
occupation numbers of different flavors, with $\nu_x$ and $\bar\nu_x$ denoting the heavy-lepton flavor of neutrinos and antineutrinos. 
In this study, as also commonly observed in state-of-the-art 
CCSN and NSM simulations,
we assume that
$\nu_x$ and $\bar\nu_x$ have similar angular distributions. The expression in Eq.~(\ref{Eq:G}) then transforms into the conventional definition of the neutrino electron lepton number, $\nu$ELN.

The occurrence of FFCs  on much shorter  scales compared to typical hydrodynamical simulations of CCSNe and NSMs, 
makes their integration into the simulations a formidable task. 
One prospective approach includes performing short scale simulations of FFCs and then extrapolating the insights gained to inform broader hydrodynamic simulations~\cite{xiong2020potential,George:2020veu,Li2021a,Just:2022flt,Fernandez:2022yyv,Ehring:2023abs, Ehring:2023lcd}.
Given this, there has been a body of research on the 
assessment of FFC outcomes  in local dynamical simulations with periodic boundary conditions~\cite{Bhattacharyya:2020dhu,Bhattacharyya:2020jpj,Wu:2021uvt,Richers:2021nbx,Zaizen:2021wwl,Richers:2021xtf,Bhattacharyya:2022eed,Grohs:2022fyq,Abbar:2021lmm,Richers:2022bkd,Zaizen:2022cik,Xiong:2023vcm},
where quasistationary flavor states have been observed in the neutrino gas.
In particular, it has been demonstrated that such states can be accurately described by analytical formulations~\cite{Xiong:2023vcm}.

Despite this, incorporating FFCs into CCSN and NSM simulations is still challenging. The obstacle arises from the need for complete angular distributions of neutrinos, a demanding task in computationally intensive simulations.

As an alternative, advanced simulations often simplify neutrino transport using a limited set of angular distribution moments~\cite{Shibata:2011kx, Cardall:2012at, thorne1981relativistic}. In 
a multi-energy neutrino gas, one can define the radial moments (with a focus on axisymmetric crossings) for each energy bin as,
\begin{equation}
I_{n,i} = \frac{E_{\nu,i}^2 \Delta E_{\nu,i}}{(2\pi)^3}  \int_{-1}^{1} \mathrm{d}\mu\ \mu^n\ \int_0^{2\pi} \mathrm{d} \phi_\nu\
        f_{\nu}(\mathbf{p}),
\end{equation}
with $E_{\nu,i}$ and $\Delta E_{\nu,i}$  being the mean energy and the width of the $i$-th energy bin, where $I_{0,i} = n_{\nu,i}$ is the number of neutrinos in that specific bin. Following 
these moments allows for a computationally more manageable treatment of the neutrino transport.  In practice, simulations typically offer only $I_{0,i}$ and $I_{1,i}$. The challenge is then to determine the ultimate values of $I_{0,i}$ and $I_{1,i}$ following FFCs. Note that the energy-integrated $I$'s can be simply obtained by a summation over $I_i$'s.

Recently, we showed that  the asymptotic outcomes of FFCs in a single-energy neutrino gas  in the moments scenario can be successfully predicted by using artificial neural networks (NNs)~\cite{Abbar:2023ltx}. In particular, 
we employed physics-informed neural networks (PINNs), where the learning and  performance of the NN   can be enhanced with the utilization of the domain knowledge~\cite{karniadakis2021physics,cuomo2022scientific, raissi2019physics}.
Our findings demonstrated the efficacy of a single hidden layer PINN, achieving a
remarkable accuracy for the prediction of the asymptotic values of $I_0$ and $I_1$ in a single-energy neutrino gas.

In this paper, we extend our prior research  by considering a multi-energy neutrino gas, which is considered a more realistic scenario. We  demonstrate that  when the FFC growth rate surpasses that of the vacuum Hamiltonian significantly, all neutrinos, irrespective of energy, share a common survival probability as dictated by the energy-integrated neutrino spectrum, consistent with the findings of Ref.~\cite{Martin:2021xyl}.

To predict the asymptotic outcome of FFCs, we employ a PINN. This PINN utilizes critical information derived from the initial (anti)neutrino zeroth and first moments, considering both the energy-integrated neutrino spectra and a specific neutrino energy bin. Consequently, it produces the corresponding moments specific to the asymptotic outcome of FFCs for that energy bin.
Our findings highlight the effectiveness of a single hidden layer PINN, achieving remarkable accuracy in predicting the asymptotic values of $I_0$ and $I_1$ for each neutrino energy bin.

This paper is organized as follows. In Sec.~\ref{sec:FFC}, we begin by providing an overview of our simulations concerning FFCs in a multi-energy neutrino gas. We also elaborate on the assumptions made in deriving the outcomes of FFCs. Moving to Sec.~\ref{sec:PINN}, we describe the architecture of our NNs, elaborating on the necessary feature engineering and the implementation of a tailored loss function. Furthermore, we present and discuss our results in this section. Finally, our findings are summarized, and conclusions are presented in Sec.~\ref{sec:dis}.

  \section{FFCs in a multi-energy neutrino gas}\label{sec:FFC}
  
  In this section, we present the results of our simulations of FFCs in a multi-energy neutrino gas. 
Essentially, when the growth rate of FFCs, $\kappa$, significantly surpasses the vacuum frequency, $\omega$, i.e. $\kappa \gg \omega$, one anticipates that neutrino energy becomes inconsequential to their flavor evolution. 
Here, $\omega\equiv \delta m^2/(4E_\nu)$ with $\delta m^2$ being the squared neutrino mass difference.
This suggests that in such circumstances, all neutrinos should experience identical survival probabilities dictated solely by the energy-integrated neutrino spectrum, effectively making the energy irrelevant.

The condition $\kappa \gg \omega$ could be expected to be met in a dense neutrino gas provided that $\lambda \gg \omega$~\footnote{Note that this is not true if one considers  
 an oversimplified  model
(that constraints the  activation of inhomogeneous 
 unstable modes),
 or if the $\nu$ELN crossing is too narrow/shallow. In such situations, the growth rate of FFCs could be significantly suppressed so that $\kappa \gg \omega$ does not hold anymore~\cite{DedinNeto:2023ykt}.}, where $\lambda=\sqrt{2}G_{\rm F}n_{\nu_e}$ with $n_{\nu_e}$ being here the initial $\nu_e$ number density of the neutrino gas. 
 A crucial exception arises when 
 the neutrino gas lepton asymmetry ratio
  defined as,
\begin{equation}
\alpha=n_{\bar\nu_e}/n_{\nu_e},    
\end{equation}
is extremely close to unity, which already implies that flavor equipartition should occur on short scales in the neutrino gas, even in the absence of FFCs (see the discussion in Ref.~\cite{abbar2020fast}). 

In the following, we first demonstrate that when $\lambda \gg \omega$, all neutrinos (with different energies) experience identical survival probabilities. Subsequently, we discuss  an analytical representation of the survival probabilities, which will be useful for our PINN calculations.  
  
\subsection{Results of the simulations}\label{sec:P_s}  

We consider a multi-energy and multi-angle neutrino gas in a 1D box, extending the framework outlined in Ref.~\cite{Wu:2021uvt}.
Our model assumes translation symmetry along the $x$ and $y$ axes, axial symmetry around the $z$ axis, and employs periodic boundary conditions in the $z$ direction. 
We also take two flavor approximation, exclude the consideration of neutrino-matter forward scattering, and assume that the system consists of (anti)neutrinos of electron flavor whose energy and angular distribution is spatially homogeneous in the beginning for simplicity.
Under these assumptions, the evolution of the normalized neutrino and antineutrino density matrices, $\varrho(t,z,\omega,\mu)$ and $\bar\varrho(t,z,\omega,\mu)$ is governed by the following equations of motion, 

\begin{align}\label{eq:eom}
& (\partial_t+ \mu\partial_z)\varrho(t,z,\omega,\mu)
=-i[H(t,z,\omega,\mu),\varrho(t,z,\omega,\mu)],\nonumber\\
& (\partial_t+ \mu\partial_z)\bar\varrho(t,z,\omega,\mu)
=-i[\bar H(t,z,\omega,\mu),\bar\varrho(t,z,\omega,\mu)],
\end{align}
where  $\mu$ here represents the neutrino velocity in the $z$ direction. 
The Hamiltonian $H(t,z,\omega,\mu)$ and $\bar H(t,z,\omega,\mu)$ are given by $H(t,z,\omega,\mu)=H_{\rm vac}(\omega)+H_{\nu\nu}(t,z,\mu)$ 
and $\bar H(t,z,\omega,\mu)=H_{\rm vac}(\omega)-H^*_{\nu\nu}(t,z,\mu)$, where
\begin{equation}\label{eq:hvac}
H_{\rm vac}=\omega
\left [
\begin{array}{cc}
-\cos2\theta_{\rm eff} & \sin2\theta_{\rm eff}\\
\sin2\theta_{\rm eff} & \cos2\theta_{\rm eff} \\
\end{array}
\right ],
\end{equation}
with $\theta_{\rm eff}$ the effective vacuum mixing angle, and
\begin{align}\label{eq:hnu}
 H_{\nu\nu}&(t,z,\mu)= 
~ \lambda \int_{-1}^1 \mathrm{d}\mu' \int_0^\infty \mathrm{d}\omega'(1-\mu \mu')\times\\
& \left[  f_\nu(\omega',\mu')\varrho(t,z,\omega',\mu')-
\alpha f_{\bar\nu}(\omega',\mu')\bar\varrho^*(t,z,\omega',\mu')\right].\nonumber
\end{align}
Here, the neutrino distribution functions are
similar to those introduced in Eq.~(\ref{Eq:G}) except that
they are integrated over $\phi_\nu$ and now normalized by $\int \mathrm{d}\mu \mathrm{d}\omega f_{\nu(\bar\nu)}(\omega,\mu)=1$.  

For the specific simulation discussed below, we take a 1D box of size $L=1200\lambda^{-1}$ with $\alpha=0.9$. 
The neutrino distribution function is parameterized by 
\begin{equation}
f_\nu(\omega,\mu) \propto
\omega^{-(\chi_\nu+2)}
\exp{\left[-\frac{(\mu-1)^2}{2\sigma_{\nu}^2}-\frac{(1+\chi_\nu)\delta m^2}{4\omega\langle E_\nu\rangle}\right]},
\end{equation}
with $\sigma_\nu=0.6$, $\sigma_{\bar\nu}=0.5$, $\chi_\nu=3.2$, $\chi_{\bar\nu}=4.5$, $\langle E_\nu\rangle=10$~MeV, and $\langle E_{\bar\nu}\rangle=12$~MeV. 
For the vacuum mixing parameters, we set $\omega/\lambda=10^{-4}$ for $E_\nu=1$~MeV with $\theta_{\rm eff}=10^{-5}$.
In order to introduce small inhomogeneity to the system, we follow Ref.~\cite{Wu:2021uvt} to assign perturbations to $\varrho$ and $\bar\varrho$ at $t=0$ by 
\begin{align}
 & \varrho_{ee}(z,\mu) = \bar\varrho_{ee}(z,\mu) = (1+\sqrt{1-\epsilon^2 (z)})/2, \nonumber \\
 & \varrho_{xx}(z,\mu) = \bar\varrho_{xx}(z,\mu) =(1-\sqrt{1-\epsilon^2 (z)})/2, \nonumber \\
 & \varrho_{ex}(z,\mu)  =\bar\varrho_{ex}(z,\mu) = \epsilon (z)/2,
\end{align}
where $\epsilon(z)$ is a real number randomly generated between $0$ and $0.01$.
We discretize the simulation domain with 6000, 50, and 20 uniform grids in $-600\leq z\leq 600$~$\lambda^{-1}$, $-1\leq \mu\leq 1$, and $0\leq 4 \omega/\delta m^2\leq 0.5$~MeV$^{-1}$, and use the finite difference scheme of \textsc{cose}$\nu$~\cite{george2023cosenu} to conduct the simulation until $t=2000\lambda^{-1}$ when the system has settled into the asymptotic state.

\begin{figure} [tb]
\centering
\begin{center}
\includegraphics*[width=.45\textwidth]{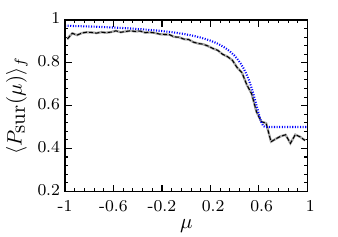}
\end{center}
\caption{
Comparison of the spanned range of spatially averaged survival probabilities for neutrinos with different $\omega$ (grey shaded area; see Eq.~\eqref{eq:Psur-z}), with the survival probability averaged over space and $\omega$ (black dashed curve; see Eq.~\eqref{eq:Psur-zw}), at the final time of the simulation for a system with $\omega\ll \lambda$. 
Also shown is the analytical prescription for $\langle P_{\rm sur} (\mu,\omega) \rangle_f$ (blue dotted curve) described in Eq.~(\ref{eq:sur2f}).
Note that  the grey-shaded area basically
overlaps with the black dashed line, implying that the survival probabilities are nearly independent of the neutrino energy.
}
\label{fig:simPee}
\end{figure}

Fig.~\ref{fig:simPee} shows the survival probability of electron neutrinos as a function of $\mu$,  averaged over $z$ and $\omega$ (black dashed curve) at the end of the simulation, computed as,
\begin{align}\label{eq:Psur-zw}
    \langle P_{\rm sur} (\mu) \rangle_f =\int \mathrm{d}z \mathrm{d}\omega f_\nu(\mu,\omega) \varrho_{ee}\Big/\int \mathrm{d}z \mathrm{d}\omega f_\nu(\mu,\omega),  
\end{align}
as well as the spanned range of the spatially averaged survival probabilities (shaded grey area) by all different $\omega$ values calculated by:
\begin{align}\label{eq:Psur-z}
    \langle P_{\rm sur} (\mu,\omega) \rangle_f =\int \mathrm{d}z f_\nu(\mu,\omega) \varrho_{ee}\Big/\int \mathrm{d}z f_\nu(\mu,\omega). 
\end{align}

This comparison clearly shows that $\langle P_{\rm sur} (\mu,\omega) \rangle_f$ is nearly independent of $\omega$ as the grey-shaded area basically overlaps with the black dashed line.
Also shown in the plot is the analytical prescription for $\langle P_{\rm sur} (\mu,\omega) \rangle_f$ (blue dotted curve) following Ref.~\cite{Xiong:2023vcm} for two-flavor scenario described in Eq.~(\ref{eq:sur2f}).

While  $\omega/\lambda \lesssim 10^{-4}$ is a reasonable assumption regarding  the SN neutrino decoupling region, we conducted additional calculations with $\omega/\lambda=10^{-3}$. Although the
 spanned range of spatially averaged survival probabilities turned out to be 
 more noticeable and the analytical formula was less precise in that case (compared to the former case), we noticed that assuming an energy-independent survival probability remains a justified assumption also for $\omega/\lambda \gtrsim 10^{-3}$.

\subsection{Survival probability function}\label{sec:P_s}  
To effectively train and evaluate our PINN, we require the survival probabilities derived from energy-integrated neutrino distributions. Our approach involves utilizing two parametric distributions for the initial neutrino angular distributions, as previously explored in our work~\cite{Abbar:2023ltx}: the maximum entropy distribution and the Gaussian distributions~\cite{Cernohorsky:1994yg, Richers:2022dqa, abbar2023applications,Abbar:2023zkm, Wu:2021uvt, Yi:2019hrp}, defined as,
\begin{equation}\label{eq:fmu}
\begin{split}
f^{\rm{max-ent}}_\nu(\mu) &= A\exp(a\mu),\\
f^{\rm{Gauss}}_\nu(\mu) &= A\exp\big(-\frac{(1-\mu)^2}{\xi}\big),
\end{split}
\end{equation}
respectively, where,
\begin{equation}
 f_{\nu}(\mu) =  \int_0^\infty \int_0^{2\pi} \frac{E_\nu^2 \mathrm{d} E_\nu \mathrm{d} \phi_\nu}{(2\pi)^3} 
        f_{\nu}(\mathbf{p}).
\end{equation}
Note that here $A$, $a$ and $\xi$ are arbitrary parameters  which determine the overall neutrino number and  the shape of the distributions. Allowing for two distinct forms of angular distributions takes into consideration potential deviations in the shape of neutrino angular distributions in realistic simulations, which can occur, e.g., due to the use of different closure relations.

In our analytical treatment of the survival probability, we follow closely our recent works in Refs.~\cite{Xiong:2023vcm, Abbar:2023ltx}. 
We assume that 
 $G(\mu)\ (=\int_{0}^{2\pi}\mathrm{d}\phi_\nu G(\mathbf{v}))$ has only one zero crossing.
In the three-flavor scenario, the survival probability can then be defined as,
 \begin{equation}\label{eq:sur}
    P_{\mathrm{sur}}(\mu) = 
    \begin{cases}
        \frac{1}{3} & {\rm for~}\mu^<, \\
        \mathcal{S}_3(\mu) & {\rm for~}\mu^>,
    \end{cases}
\end{equation}
with,
\begin{equation}\label{eq:express_continuous}
    \mathcal{S}_3(\mu) = 1-\frac{2}{3}h(|\mu-\mu_c|/\zeta).
\end{equation}
where $h(x) = (x^2+1)^{-1/2}$ and $\zeta$ can be found such that the survival
probability function is continuous. Here, $\mu^{<} (\mu^{>})$ are defined as the $\mu$ range
over which the following integral is smaller (larger):
  \begin{equation} 
 \begin{split}
 \Gamma_{+}&=\bigg| \int_{-1}^{1} \mathrm{d}\mu\ G(\mu) \Theta[G(\mu)] \bigg|, \\
 \Gamma_{-}&=\bigg| \int_{-1}^{1} \mathrm{d}\mu\ G(\mu) \Theta[-G(\mu)] \bigg|,
 \end{split}
 \end{equation} 
where $\Theta$ is the Heaviside theta function. 
In the case of the two-flavor scenario, the survival probability can be obtained  using:
\begin{equation}\label{eq:sur2f}
    P_{\mathrm{sur}}(\mu) = 
    \begin{cases}
        \frac{1}{2} & {\rm for~}\mu^<, \\
        \mathcal{S}_2(\mu) & {\rm for~}\mu^>,
    \end{cases}
\end{equation}
with,
\begin{equation}\label{eq:express_continuous}
    \mathcal{S}_2(\mu) = 1-\frac{1}{2}h(|\mu-\mu_c|/\zeta).
\end{equation}
We refer
an interested reader to Refs.~\cite{Xiong:2023vcm} for more details.
  
 Using the neutrino angular distributions in Eq.~(\ref{eq:fmu}) and the survival probability function defined in Eq.~(\ref{eq:sur}), one can obtain  the asymptotic outcomes of FFCs given the initial distributions.

\section{Applications of Neural Networks}\label{sec:PINN}

To effectively train our PINN, we require information on two fronts: the energy-integrated moments of the neutrino gas and the  moments within a specific energy bin. The former implicitly contains the necessary information for the survival probability (dictated only by the energy-integrated quantities), while the latter supplies the bin-specific information to which  the survival probability must be applied.

To prepare our datasets, we begin with the initial energy-integrated angular distributions of neutrinos, which can follow either a maximum entropy or a Gaussian distribution. With these distributions for $\nu_e$ and $\bar\nu_e$, we derive analytical survival probabilities.
Next, we apply these analytical distributions to the neutrino angular distributions within a particular energy bin (again either maximum entropy  or a Gaussian). This process helps us determine the eventual outcomes of FFCs for that specific bin. By performing integration over the neutrino angular distributions, we can then obtain the initial and final values of $I_0$'s and $I_1$'s for that specific energy bin.

Before discussing our findings, it's crucial to emphasize  that to
 ensure high performance in our NN models on the test set, it's essential to divide the dataset into three distinct sets: Training set for foundational learning, development set for optimizing hyper-parameters, and test set for evaluating the model's generalization to novel data.

\subsection{The architecture of NNs}

For a given multi-energy neutrino gas, one is provided with the initial values of energy-integrated $I_0$'s and $I_1$'s of $\nu_e$, $\bar\nu_e$ (also of $\nu_x$, which is irrelevant here since it has no effect on  the survival probability). 
In addition, for each specific energy bin, one has $I_0$'s and $I_1$'s of $\nu_e$, $\bar\nu_e$, and $\nu_x$.
In this context, we make the assumption that the initial distributions of $\bar\nu_x$ and $\nu_x$ are identical (though their final ones following FFCs could be different), a simplification that aligns with the majority of state-of-the-art CCSN and NSM simulations.

Though in total 10 $I$'s  are available (which could be, in principle, the inputs of  NNs), we here introduce a layer of feature engineering
 to enhance the performance of our NNs, namely we define the new features:
\begin{equation}\label{eq:inputs}
\alpha, F_{\nu_e}, F_{\bar\nu_e}, n_{\nu_e,i}, n_{\bar\nu_e,i},  n_{\nu_x,i}, F_{\nu_e,i}, F_{\bar\nu_e,i}, F_{\nu_x,i},
\end{equation}
with  $F_{\nu} = (I_1/I_0)_{\nu}$. Here, the quantities without/with subscript $i$ indicate the quantities belonging to the energy-integrated spectrum/specific energy bin. Note that the neutrino number densities in the particular energy bin must be smaller than the corresponding energy-integrated values.

The selection of these features offers explicit insights into the configuration of neutrino angular distributions, which plays a crucial role in understanding the asymptotic outcome of FFCs. Furthermore, it is worth highlighting that all quantities in this context are normalized by the initial energy-integrated $\nu_e$ number density, allowing the convenient choice of setting it to $n^{\rm{initial}}_{\nu_e} = 1$. This simplification  reduces the number of inputs to our NNs, and notably, there is no input parameter related to $n_{\nu_e}$.

As we also discussed in our previous work~\cite{Abbar:2023ltx},
there is still the possibility of improving our NNs  through more advanced feature engineering.
By considering the neutrino survival probability's shape, as expressed in Eq.~(\ref{eq:sur}), one observes that 
a significant amount of information about the shape of the survival probability can be derived by learning the position of $\mu_c$. 
Another crucial piece of information, given $\mu_c$, is determining the side where equipartition happens. The behavior of the survival probability on the opposite side is regulated by conservation laws. This side's determination is described by the quantity $E_{RL}$, a binary value that equals 1 if equipartition happens for $\mu_c\leq\mu$ and 0 otherwise.

\begin{figure} [tb]
\centering
\begin{center}
\includegraphics*[width=.45\textwidth, trim= 0 0 0 0, clip]{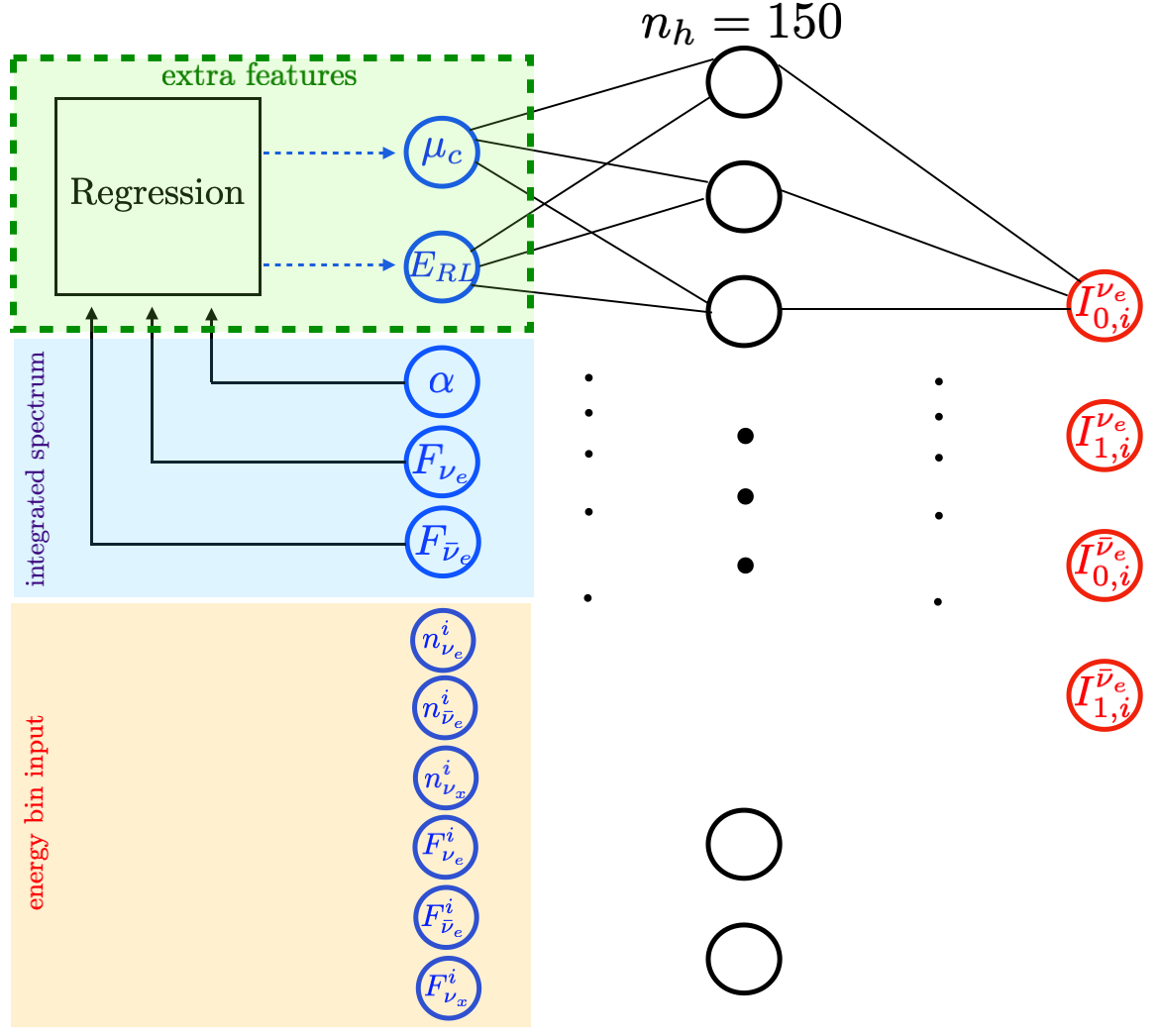}
\end{center}
\caption{
Schematic architecture of our NNs. The green zone shows the implementation of the extra features, $\mu_c$,
and $E_{RL}$, which are obtained through an extra layer of regression, using  linear and logistic regressions, respectively. Here, $\mu_c$ is the
crossing direction and $E_{RL}$ is a binary, which is 1 if the equilibrium occurs for $\mu_c\leq\mu$,
and 0 otherwise. 
The blue zone represents energy-integrated inputs, while the orange zone displays inputs for specific energy bins.
Note that the neutrino number densities in the particular energy bin must be smaller than the corresponding energy-integrated values.
In our basic NN, referred to as the NN with no extra features, the NN only takes the inputs highlighted in Eq.~(\ref{eq:inputs}).
However and in our PINN, we provide our NN with the extra features $\mu_c$  and $E_{RL}$.
}
\label{fig:scheme}
\end{figure}

As in Ref.~\cite{Abbar:2023ltx} and as illustrated in Fig.~\ref{fig:scheme},  we explore two distinct architectures in our NN framework.
In the foundational architecture, we integrate only $\alpha, F_{\nu_e},  F_{\bar\nu_e}, n_{\nu_e,i}, n_{\bar\nu_e,i},  n_{\nu_x,i}, F_{\nu_e,i}, F_{\bar\nu_e,i}, \mathrm{and}\ F_{\nu_x,i}$ into our NN. 
On the other hand, our  alternative NN  includes also information coming from $\mu_c$ and $E_{RL}$.
Our feedforward NN has a single hidden layer containing 150 neurons, 
as justified in Fig.~\ref{fig:num_hid} and the text around it.

Regarding the output layer, our NNs return $I_{0,i}$ and $I_{1,i}$ for $\nu_e$ and $\bar\nu_e$, employing a total of 4 neurons. Deriving $I_{i}$'s  for $\nu_x$ and $\bar\nu_x$ is achieved by applying principles of neutrino and antineutrino number density, as well as momentum conservation. Put simply, our NN ensures neutrino conservation laws.


Apart from the inputs, we also consider modifying the loss function of our NN's.
In particular, we introduce  an additional loss term in the optimization of the NN model with the extra features, defined as,
\begin{equation}
   \mathcal{L}_{\rm{extra}} = \frac{1}{N_{\rm{sample}}} \Sigma_k (\Delta N_{{\nu_e} + {\bar\nu_e}, k})^2,  
 \end{equation}
 which tends to penalize any deviation in the  number of neutrinos in the electron channel, i.e., $ N_{{\nu_e} + {\bar\nu_e}} = n_{\nu_e,i} + n_{\bar\nu_e,i}$, a critical parameter of utmost significance in CCSNe and NSMs.
Here  $\Delta$, $N_{\rm{sample}}$, and $\Sigma_k$ denote the difference between the true and predicted values, the number of samples in the training set, and the summation over the training samples, respectively.

The specific inclusion of the domain knowledge allows one to consider this particular NN architecture as a PINN~\cite{karniadakis2021physics,cuomo2022scientific, raissi2019physics}. 
 The PINN should be compared with our basic NN, referred to as NN with \emph{no extra features}, for which the loss term only includes the ordinary mean squared errors of the output parameters.
 

 \subsection{The NN's performance}

\begin{figure} [tb!]
\centering
\begin{center}
\includegraphics*[width=0.46\textwidth, trim= 5 0 0 0, clip]{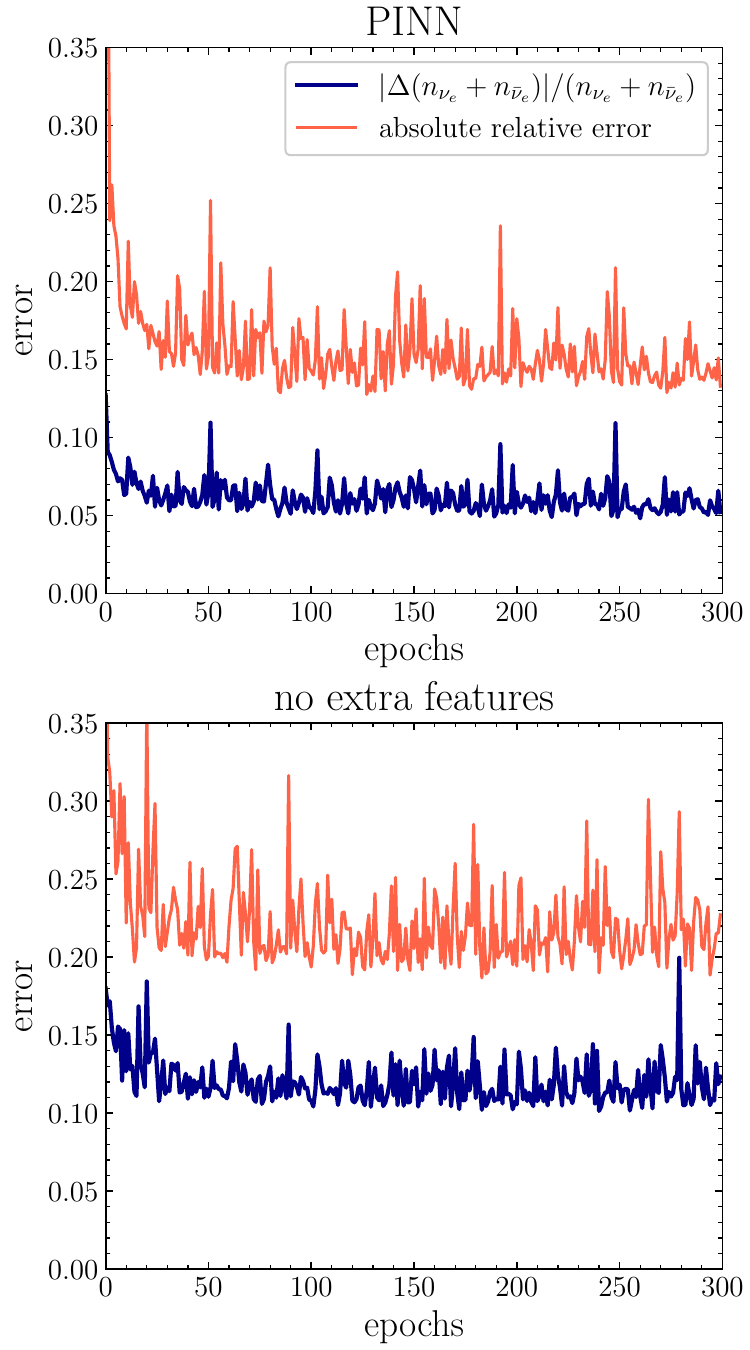}
\end{center}
\caption{
Performance evaluation of the PINN and the basic NN with no extra features.
We present the relative absolute error in the output parameters, along with the relative error in the total number of neutrinos within the electron channel,  $N_{{\nu_e} + {\bar\nu_e}}$. 
It is evident that the PINN  can well outperform the basic NN with no extra features.
Here, an epoch  refers to a single pass through the entire training dataset during the training phase. 
}
\label{fig:error}
\end{figure}

In this section,
we present and discuss the performance of our NNs in predicting the  asymptotic outcome of FFCs  in a multi-energy three-flavor neutrino gas.
For training/testing our NNs, we generate  a dataset comprising a well-balanced combination of maximum entropy and Gaussian initial neutrino angular distributions.
The ultimate outcome of FFCs is determined through a three-flavor survival probability, as detailed in Eq.~(\ref{eq:sur}).
We also set $\alpha \in (0,2.5)$, $F_{\nu_x, (i)} \in (0,1)$,  $F_{\bar\nu_e, (i)} \in (0.4F_{\nu_x, (i)}, F_{\nu_x, (i)})$, and $F_{\nu_e, (i)} \in (0.4 F_{\bar\nu_e, (i)}, F_{\bar\nu_e, (i)})$, which is consistent with the expected hierarchy $F_{\nu_e} \lesssim F_{\bar\nu_e} \lesssim F_{\nu_x}$. Regarding $n_{\nu, i}$'s, we take them from a half-normal distribution with zero mean and a standard deviation of $0.1 n_\nu$, with $n_\nu$ being the energy-integrated neutrino number density.
This choice can enhance the performance of our NNs in the energy bins with fewer neutrinos.
Also note that since our NNs process only
 a single energy bin at a time,
the hierarchy of flux factors among different neutrino energies is irrelevant here.

In Fig.~\ref{fig:error}, we illustrate the performance of our PINN and the basic NN without extra features. The relative error in the electron neutrino number density within our PINN, quantified by $|\Delta (n_{\nu_e, i} + n_{\bar\nu_e, i})|/(n_{\nu_e, i} + n_{\bar\nu_e, i})$, achieves a minimum of 6\%. Additionally, the mean absolute relative error in the output variables, computed as the mean of $ |\Delta I_i |/I_i$, attains values  $\sim 16\%$.
In contrast, when considering the basic NN, these errors increase to  $\sim 12\%$ and $\sim 22\%$, respectively, showing  higher discrepancies. The noticeable performance improvement within our PINN can be primarily attributed to the  inclusion of extra features, which provide extra information on the shape of the survival probability distribution.

Comparing the findings illustrated in Fig.~\ref{fig:error} with those discussed in Ref.~\cite{Abbar:2023ltx}, a discerning reader will observe a substantial discrepancy in the impact of employing PINN. Specifically, the application of PINN results in a significantly more pronounced enhancement in performance in the former case. While the utilization of PINN can almost reduce the error by a factor of two in the multi-energy neutrino gas, its application to a single-energy neutrino gas only yields a modest $\lesssim 1\%$ improvement in the error.

This discrepancy can be attributed to the substantial difference in the amount of input information between the two cases. In the former scenario, the volume of input information is notably larger, leading to a higher degree of degeneracy in the input data. The introduction of PINN in this context is remarkably effective in mitigating this degeneracy and, consequently, substantially reducing the error.

\begin{figure} [tb!]
\centering
\begin{center}
\includegraphics*[width=.46\textwidth, trim= 5 0 0 0, clip]{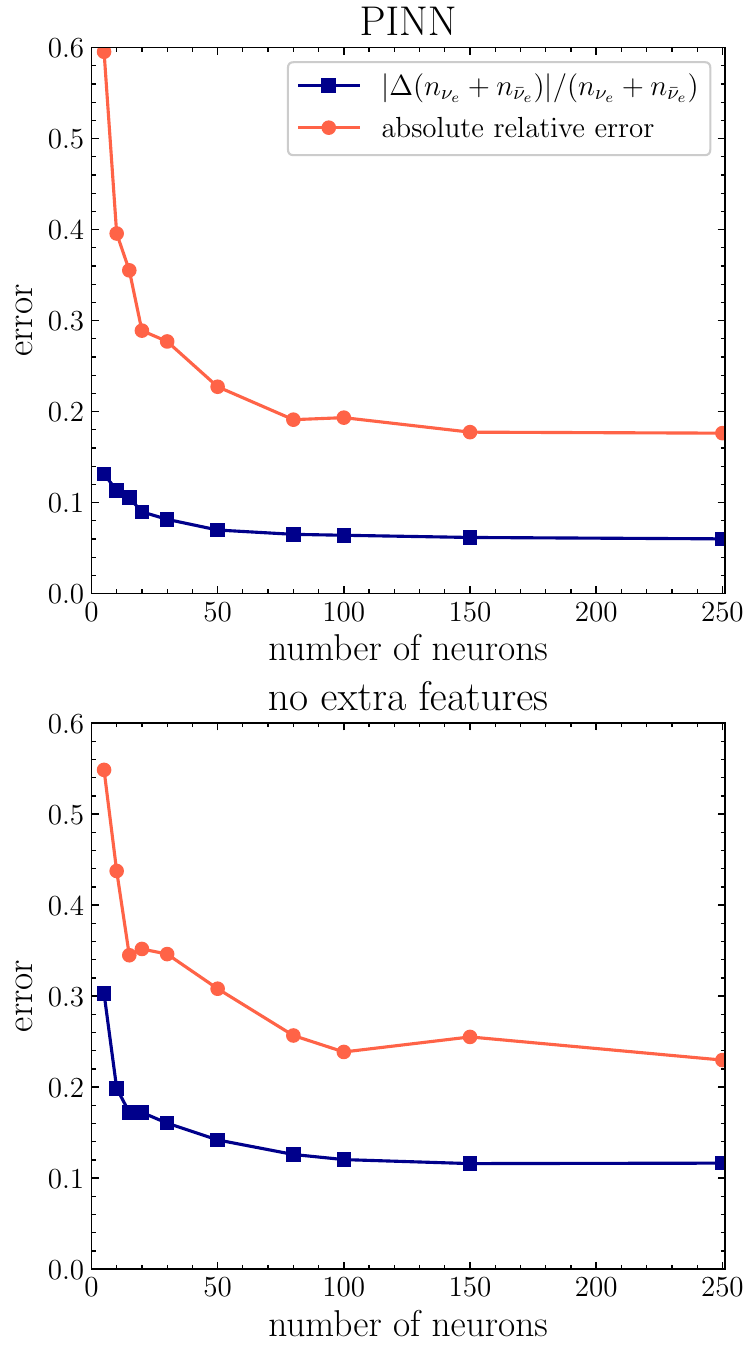}
\end{center}
\caption{
Performance evaluation of our PINN and the basic NN with no extra features (on the validation set) 
 as a function of the number of neurons in the hidden layer.
It is evident that the NNs  have achieved their  best performance on the validation set once $n_h\gtrsim 150$.
The labels and NN models are the same as those in Fig.~\ref{fig:error}.
}
\label{fig:num_hid}
\end{figure}

The computations conducted here have utilized a feedforward neural network featuring a single hidden layer having $n_h = 150$ neurons. The reasoning behind selecting this specific number of neurons is depicted in Fig.~\ref{fig:num_hid}, which shows  errors for different NN architectures. The optimal performance on the validation set is observed when $n_h\gtrsim 50$.

\begin{figure} [tb!]
\centering
\begin{center}
\includegraphics*[width=.46\textwidth, trim= 0 0 35 10, clip]{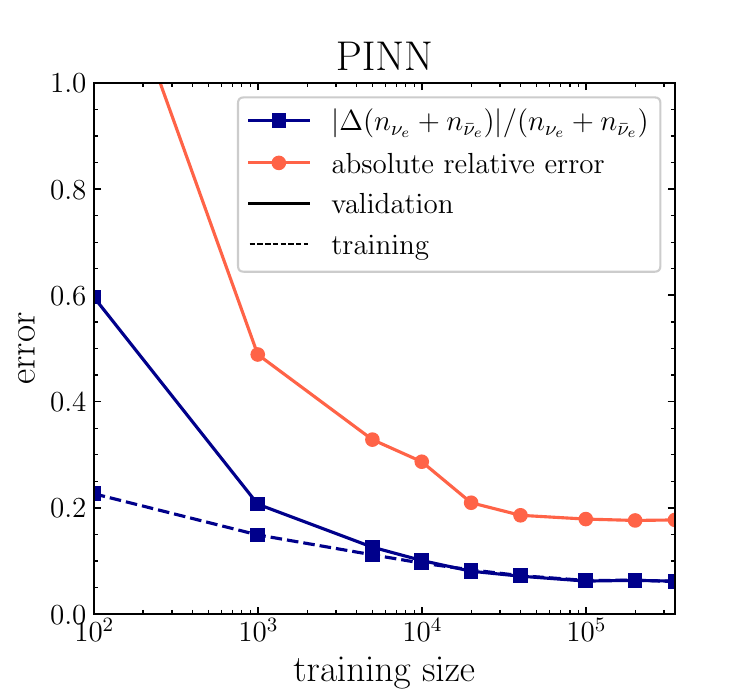}
\end{center}
\caption{
Absolute relative error in the output of our PINN (red curve) vs the relative error in the number of neutrinos
in electron channel, i.e., $N_{{\nu_e} + {\bar\nu_e}}$ (blue curve).
 The inclusion of a few thousand
data points in the training set leads to the disappearance of error variations between the validation and training sets.
Note that we do not display the absolute relative error in the training set. This is due to the presence of small $I's$ in the training set (which are removed from the test set), causing a significantly larger relative error. Hence, any direct comparisons between the absolute relative errors in the training and test sets would be unfair.
}
\label{fig:PINN_error}
\end{figure}

In Fig.~\ref{fig:PINN_error}, we analyze our PINN's performance concerning the training set size. The red curve indicates the absolute relative error in the PINN's output, while the blue curve shows the relative error in $N_{{\nu_e} + {\bar\nu_e}}$. As the training dataset expands to include several thousand data points, the errors rapidly decrease to approximately $\sim 6\%$ and $\sim 18\%$, respectively, and additionally, the disparity between the validation and training set errors diminishes. These findings align with results observed in single-energy scenario calculations~\cite{Abbar:2023ltx}. This underscores the crucial minimum number of data points required for reliable calculations using NNs.

There remains a crucial aspect regarding the assessment of the absolute relative error that needs discussion. In our prior study concerning FFCs within a single-energy neutrino gas~\cite{Abbar:2023ltx}, we primarily regarded the absolute error as the relevant metric.
However, when addressing the complexities of a multi-energy neutrino gas, the absolute error falls short.
The problem arises because, despite normalizing all quantities by $n_{\nu_e}$, the values of $I$'s
within a specific energy bin are expected to constitute a minor fraction of one. Hence, achieving a low absolute error does not inherently guarantee accurate prediction due to the relatively small magnitudes involved.

Hence and for the multi-energy neutrino gas, we've adopted the absolute relative error as the informative metric. Nonetheless, this choice comes with a notable drawback: extremely small $I$'s  yield disproportionately large absolute relative errors. While these cases may not include the most intriguing aspects of the parameter space, their associated errors tend to dominate over the rest of the parameter space.
To resolve this issue, in our performance evaluation on the test set, we have chosen to exclude data points where $|I_i| \leq 5\times10^{-3}$ (while retaining them in the training set). 
In Sec.\ref{sec:spec}, we come back to this problem
 and devise a solution to this challenge.

 \subsection{Reconstruction of  neutrino energy spectra}\label{sec:spec}

\begin{figure*} [tb!]
\centering
\begin{center}
\includegraphics*[width=1.05\textwidth, trim= 0 0 0 0, clip]{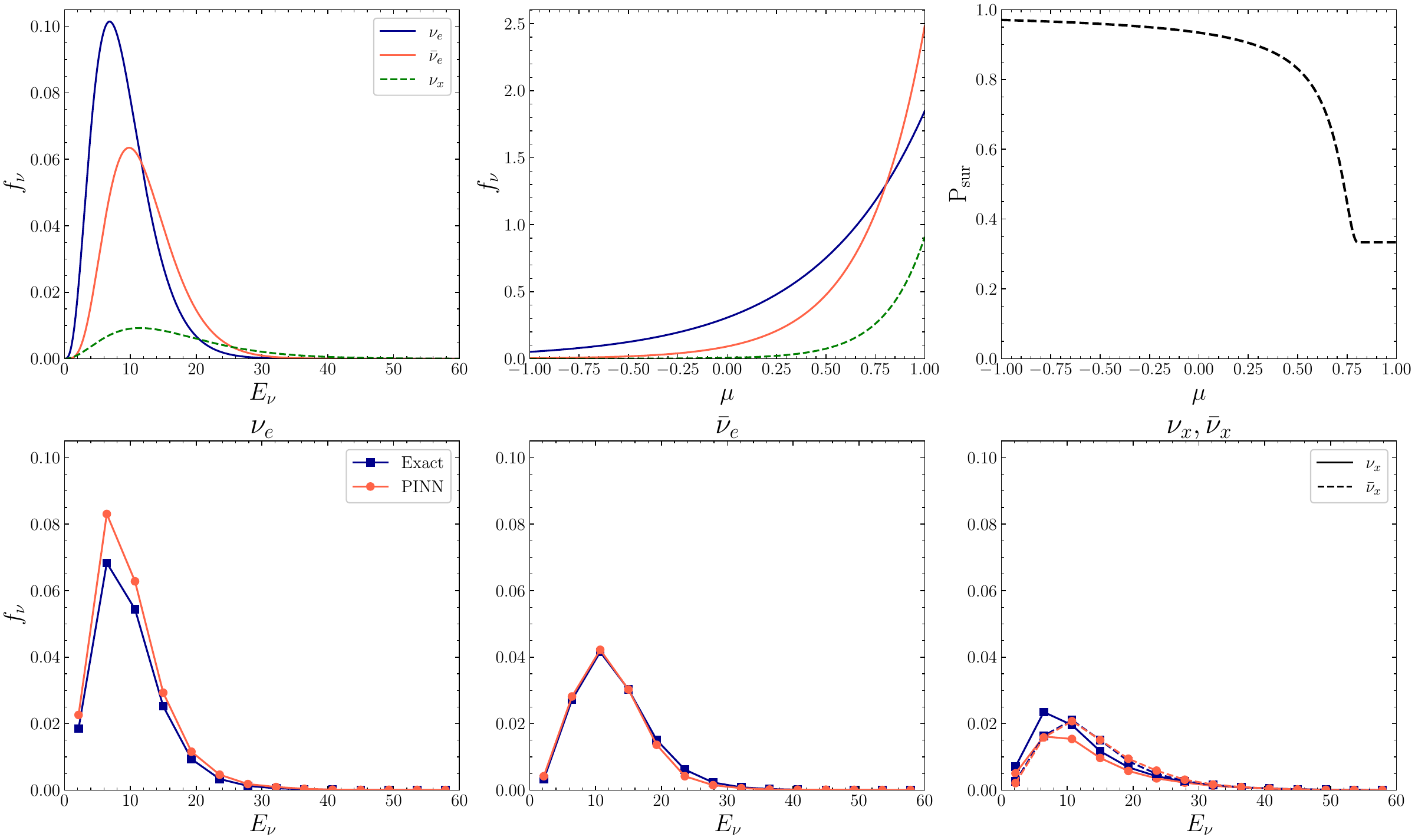}
\end{center}
\caption{
Performance evaluation of our PINNs in reconstructing neutrino energy spectra. The upper panels exhibit the initial spectra characteristics, including the neutrino initial energy spectra, angular distributions of the energy-integrated neutrino spectra as a function of $\mu$, and the corresponding FFC survival probability. Here, we have assumed $F_{\nu_e} = 0.5$, $F_{\bar\nu_e} = 0.7$, and $F_{\bar\nu_e} = 0.8$. The lower panels illustrate the post-FFCs final neutrino spectra for $\nu_e$, $\bar\nu_e$, and $\nu_x$ ($\bar\nu_x$), respectively. The results are shown for both the PINN approach and an exact method, assuming having access to the full neutrino angular distributions.
As one can see, the prediction errors for antineutrinos are much smaller than those concerning neutrinos.}
\label{fig:spec}
\end{figure*}

In the preceding part, we engineered our NNs to process the energy-integrated $I$'s  and their  values within an specific bin, $I_i$'s. 
To elaborate, our methodology works on an energy bin-based approach, focusing solely on predicting the final outcomes within individual energy bins. This  approach eliminates unnecessary complexity associated with attempting to reconstruct the entire neutrino energy spectra following FFCs  at once.

In this  part, we explore the performance of our PINN's to reconstruct the complete neutrino energy spectra following FFCs.
While this application of our NNs might appear straightforward initially, it presents significant challenges.
Specifically, when our PINNs are employed to analyze the tail of the energy spectrum, where the count of neutrinos is notably low, we encounter a considerable obstacle. The relative error in these instances might surpass the total values of $n_{\nu(\bar\nu)}$, potentially leading to a scenario where the conservation laws cannot be satisfied, 
as discussed in the following.
In the high-energy tail of the spectra, where neutrino number densities can reach very low values, the predicted value of $n_{\nu_e(\bar\nu_e)}$ may even surpass the total (anti)neutrino number density. This is attributed to the expected large relative errors associated with the output of neural networks when dealing with such small values, as discussed before. 
Consequently, while the conservation law for neutrino number remains mathematically valid in principle, it loses its practical significance. This is because adhering to the conservation laws would now imply a negative number density for $n_{\nu_x(\bar\nu_x)}$, which is not physically meaningful.

To address this challenge, we devised a  solution by simultaneously adjusting the number of $\nu_e$ ($\bar\nu_e$) and $\nu_x$ ($\bar\nu_x$) while ensuring the conservation of total neutrino numbers. Our approach indeed involves the utilization of two PINNs.
The first PINN is designed to compute $I_{0,i}$'s and $I_{1,i}$'s for $\nu_e$ and $\bar\nu_e$, following the standard architecture we discussed in previous part. Meanwhile, the second PINN shares a similar structure but deals with $\nu_x$ and $\bar\nu_x$, calculating their respective $I_{0,i}$'s and $I_{1,i}$'s. 
In practice, this implies that the outputs of such a PINN are:
\begin{equation}
I^{\nu_x}_{0,i}, I^{\nu_x}_{1,i},
I^{\bar\nu_x}_{0,i}, 
I^{\bar\nu_x}_{1,i},
\end{equation}
while the input features are not different  in the two PINNs. 
Importantly, in this PINN, the $I$'s of $\nu_e$ and $\bar\nu_e$ are then derived through neutrino conservation laws. In essence, the difference between these two PINNs lies in the information they provide—while the first PINN provides $\nu_e$ $I$'s, the second one focuses on $\nu_x$, with the retrieval of $\nu_e$ quantities governed by conservation principles.

Given these two PINNs, the final neutrino number densities for each energy bin can be calculated as:
\begin{equation}
n^{\mathrm{fin}}_{\nu_\beta (\bar\nu_\beta), i} = 
\frac{N^{\mathrm{ini}}_{\nu (\bar\nu), i}}{N^{\mathrm{pred}}_{\nu (\bar\nu), i}}\
n^{\mathrm{pred}}_{\nu_\beta (\bar\nu_\beta), i},
\end{equation}
where $n^{\mathrm{pred}}_{\nu_\beta (\bar\nu_\beta), i}$ 
represents the corresponding PINN predicted value for the neutrino species $\beta$.
Here,
 $N^{\mathrm{pred}}_{\nu (\bar\nu), i}$ 
 denotes the predicted total (anti)neutrino number density, and $N^{\mathrm{ini}}_{\nu (\bar\nu), i}$ 
 is its initial value.
 It's important to note that the prediction for each neutrino species is conducted by the relevant PINN model discussed earlier, i.e., the former PINN is employed for the electron species, while the latter is used for heavy-lepton flavors.
Note that the errors in the predictions of $n_\nu$ can now be automatically adjusted to ensure respecting the conservation laws, preventing negative number densities. Furthermore, a fair treatment is now applied to electron and heavy-lepton flavors, preventing one from becoming unreasonably small when the error in the prediction of the other is unreasonably large.

In Fig.~\ref{fig:spec}, we present the performance evaluation of our PINNs in reconstructing the neutrino energy spectra. 
The upper panels show the 
 initial neutrino  distributions, which  are prepared as follows.
We describe 
the  energy-differential number flux  of a specific neutrino species,  $\nu_\beta$, as~\cite{Mirizzi:2015eza},
\begin{equation}
\mathcal{F_{\nu_\beta}}(E_\nu) \propto \frac{L_{\nu_\beta}}{ \langle E_{\nu_\beta} \rangle} f_{\nu_\beta}(E_\nu)
\end{equation}
with
\begin{equation}
f_{\nu_\beta}(E_\nu) = \frac{1}{T_{\nu_\beta} \Gamma(1+\eta_{\nu_\beta})} \bigg( \frac{E_\nu}{T_{\nu_\beta}} \bigg)^{\eta_{\nu_\beta}}  \exp(-E_\nu/T_{\nu_\beta})
\end{equation}
being the normalized $\nu_\beta$ spectrum, where $E_\nu$ is the neutrino energy. Here, $\eta_{\nu_\beta}$ and $\langle E_{\nu_\beta} \rangle$ are 
the pinching parameter and the neutrino average energy
which describe  the normalized spectrum and $T_{\nu_\beta} = \langle E_{\nu_\beta} \rangle/(1+\eta_{\nu_\beta})$. In addition,  $L_{\nu_\beta}$ is the neutrino luminosity.
To be specific, we use the following values which could be expected during the SN accretion phase~\cite{mirizzi2016supernova}:
\begin{equation}
\begin{split}
L_{\nu_e}: L_{\bar\nu_e}: L_{\nu_x} &= 1:1:0.33, \\
\langle E_{\nu_e} \rangle: \langle E_{\bar\nu_e} \rangle: \langle E_{\nu_x} \rangle &= 9:12:16.5, \\
{\eta}_{\nu_e}: {\eta}_{\bar\nu_e}: {\eta}_{\nu_x} &= 3.2:4.5:2.3. \\
\end{split}
\end{equation}
Note that  for the average energies and luminosity's only the ratios matter.
Moreover, 
we assumed these values for the energy-integrated flux factors: $F_{\nu_e} = 0.5$, $F_{\bar\nu_e} = 0.7$, and $F_{\bar\nu_e} = 0.8$. We here take 14 energy bins and for each energy bin, we then adopted an assumption describing the relationship as $F_{\nu,i} =F_{\nu}\ (70 - E_{\nu,i})^2 /60^2$. Here we have assumed that the spectra approach zero for $E_{\nu,i} \gtrsim 60$~MeV. 
Note  that we anticipate a decrease in the flux factor as neutrino energy increases. This reduction is expected to be nonlinear, attributed to the nonlinear scaling of the neutrino scattering cross-section with matter in the SN environment. While our assumption about the energy-dependent nature of the flux factor is speculative, it aligns with the expected conditions mentioned above.

In the lower panels of Fig.~\ref{fig:spec},
we present the final neutrino energy spectra following FFCs. It is evident that
a  notable difference in the spectra reconstruction error  emerges between neutrinos and antineutrinos. This disparity can be indeed quantified by an absolute spectral relative error, defined as, 
\begin{equation}
\delta_\nu = \Sigma_i \frac{n_{\nu,i}}{n_\nu} \frac{|\Delta n_{\nu,i}|}{n_{\nu,i}},
\end{equation}
where the prediction error for $n_{\nu,i}$ is weighted by the relative distribution across energy bins. For the results presented  in the lower panels of Fig.~\ref{fig:spec}, we observed $\delta_{\nu_e} = 0.20$, $\delta_{\bar\nu_e} = 0.06$, $\delta_{\nu_x} = 0.23$, and $\delta_{\bar\nu_x} = 0.05$.  
 
Despite the clear difference between neutrinos and antineutrinos in this particular example, we've noticed
that this observation depends notably 
on the specific example and it can fluctuate across different calculations and models. This variability highlights the need for more sophisticated neural network architectures, such as Bayesian neural networks. These specialized networks offer the capability to provide uncertainty estimates for predicted quantities, addressing the intricacies and fluctuations observed in these calculations.

\section{DISCUSSION AND OUTLOOK}\label{sec:dis}

We have employed a single hidden layer Physics-Informed Neural Network (PINN) to predict the asymptotic outcome of FFCs within a three-flavor multi-energy neutrino gas. Our approach focuses on utilizing the first two moments of neutrino angular distributions, making our PINNs highly relevant to state-of-the-art CCSN and NSM simulations.
We have demonstrated that our PINNs can achieve remarkable accuracy, with errors reaching $\lesssim6\%$ for the number of neutrinos in the electron channel, and $\lesssim18\%$ for the relative absolute error in the neutrino moments.

By conducting simulations of FFCs in a 1D box with periodic boundary conditions, we first demonstrated that  in scenarios where the FFC growth rate notably exceeds that of the vacuum Hamiltonian, a uniform survival probability is experienced by  all neutrinos, regardless of their energy. This common survival probability  is solely determined by the energy-integrated neutrino spectrum.

In our PINNs, we incorporated novel features to effectively capture the shape of the expected neutrino survival probability distributions. Our improvements involve incorporating the position of the zero crossing in the distribution of $\nu$ELN, $\mu_c$, and also information about the side of $\mu_c$ where the expected equipartition occurs.
Our research demonstrates that this advanced feature engineering significantly improves the performance of our PINN. 

Moreover, we demonstrated that the variance between the training and validation sets decreases significantly with a minimum of a few thousand data points. This underscores the necessity for datasets of  (at least) this size when developing more realistic models based on simulation data in future studies.

We also highlighted a significant challenge in applying NNs to predict the whole neutrino energy spectrum. This challenge arises from the fact that predicting the tail of the spectrum may lead to an error of such magnitude that it violates the preservation of neutrino conservation laws. To address this issue, we propose the development of two separate models: one dedicated to predicting electron (anti)neutrino quantities and another for heavy-lepton flavor of
(anti)neutrino quantities, respectively. By scaling the numbers in accordance with conservation laws, we could overcome this challenge. Our demonstrated approach showed that  PINNs can accurately enough reconstruct the entire neutrino spectrum, particularly in a typical neutrino spectra scenario during the SN accretion phase.

In summary, our research highlights the effectiveness of PINNs in predicting the asymptotic outcomes of FFCs within a multi-energy neutrino gas. Nevertheless, there are crucial avenues for further exploration.
An important consideration is extending our study to encompass more realistic neutrino gases characterized by non-axisymmetric distributions, where $\nu_x$ and $\bar\nu_x$ can also exhibit dissimilar patterns. 
Such refinements will  improve the feasibility of incorporating FFCs into CCSN and NSM simulations, thereby advancing our capacity to model and predict accurately these extreme astrophysical phenomena.


\section*{Acknowledgments}
We are deeply grateful to Georg Raffelt, Gabriel Mart\'inez-Pinedo, and Oliver Just for fruitful discussions.
S.A. was supported by the German Research Foundation (DFG) through
the Collaborative Research Centre  ``Neutrinos and Dark Matter in Astro-
and Particle Physics (NDM),'' Grant SFB-1258, and under Germany’s
Excellence Strategy through the Cluster of Excellence ORIGINS
EXC-2094-390783311.
M.-R.~W.\ acknowledges supports from the National Science and Technology Council under Grant No.~111-2628-M-001-003-MY4, the Academia Sinica under Project No.~AS-CDA-109-M11, and the Physics Division, National Center for Theoretical Sciences, as well as the resource of the Academia Sinica Grid-computing Center (ASGC).
Z.X. was supported by the European Research Council (ERC) under the European Union's Horizon 2020 research and innovation programme (ERC Advanced Grant KILONOVA No.~885281).
We would also like to acknowledge the use of the following softwares: \textsc{Scikit-learn}~\cite{pedregosa2011scikit}, 
\textsc{Keras}~\cite{chollet2015keras},
\textsc{Matplotlib}~\cite{Matplotlib}, \textsc{Numpy}~\cite{Numpy}, \textsc{SciPy}~\cite{SciPy}, and \textsc{IPython}~\cite{IPython}.

\bibliographystyle{elsarticle-num}
\bibliography{Biblio}

\clearpage

\end{document}